\documentclass{mn}
\usepackage{graphicx}

\def\Fbh{F_{\rm bh}}
\def\dstab{\epsilon_{\rm disk}}
\def\acool{\alpha_{\rm cool}}
\def\ebh{\epsilon_{\rm SMBH}}
\def\arh{\alpha_{\rm reheat}}
\def\ahot{\alpha_{\rm hot}}
\def\vhot{V_{\rm hot}}
\def\hMsol{h^{-1}M_{\odot}}
\def\hmpc{h^{-1}{\rm Mpc}}
\def\kms{\hbox{km}\,\hbox{s}^{-1}}
\def\gsim{ \lower .75ex \hbox{$\sim$} \llap{\raise .27ex \hbox{$>$}} }
\def\lsim{ \lower .75ex \hbox{$\sim$} \llap{\raise .27ex \hbox{$<$}} }
\def\simprop{ \lower .75ex \hbox{$\sim$} \llap{\raise .27ex \hbox{$\propto$}} }
\def\website{{\tt http://www.icc.dur.ac.uk/}}

\begin{document}
 
\title[Breaking the hierarchy of galaxy formation]{Breaking the hierarchy of 
galaxy formation} 

\author[Bower et al.]
{
\parbox[t]{\textwidth}{
\vspace{-1.0cm}
R. G. Bower$^{1}$,
A. J. Benson$^{2}$,
R. Malbon$^{1}$,
J. C. Helly$^{1}$,
C. S. Frenk$^{1}$,
C. M. Baugh$^{1}$,
S. Cole$^{1}$,
C. G. Lacey$^{1}$ 
}
\vspace*{6pt} \\
$^{1}$Institute for Computational Cosmology, Department of Physics, 
University of Durham, South Road, Durham, DH1 3LE, UK. \\
$^{2}$Dept.\ of Physics, University of Oxford, Keeble Road, Oxford, OX1 3RH, UK.
\vspace*{-0.5cm}}

\maketitle

\begin{abstract}
Recent observations of the distant Universe suggest that much of the
stellar mass of bright galaxies was already in place at $z>1$. This
presents a challenge for models of galaxy formation because massive
halos are assembled late in the hierarchical clustering process
intrinsic to the cold dark matter (CDM) cosmology. In this paper, we
discuss a new implementation of the Durham semi-analytic model of
galaxy formation in which feedback due to active galactic nuclei (AGN)
is assumed to quench cooling flows in massive halos. This mechanism
naturally creates a break in the local galaxy luminosity function at
bright magnitudes. The model is implemented within the Millennium
N-body simulation. The accurate dark matter merger trees and large
number of realisations of the galaxy formation process enabled by this
simulation result in highly accurate statistics. After adjusting the
values of the physical parameters in the model by reference to the
properties of the local galaxy population, we investigate the
evolution of the K-band luminosity and galaxy stellar mass
functions. We calculate the volume-averaged star formation rate
density of the Universe as a function of redshift and the way in which
this is apportioned amongst galaxies of different mass. The model
robustly predicts a substantial population of massive galaxies out to
redshift $z\sim 5$ and a star formation rate density which rises at
least out to $z\sim 2$ in objects of all masses. Although
observational data on these properties have been cited as evidence for
``anti-hierarchical'' galaxy formation, we find that when AGN feedback
is taken into account, the fundamentally hierarchical CDM model
provides a very good match to these observations.

\end{abstract}

\section{Introduction}

The most basic property of the galaxy population is the number density
of galaxies of different luminosity -- the luminosity function. Simple
scaling arguments suggested that the observed break in the luminosity
function at bright magnitudes would naturally arise from the long
cooling time of gas in large halos (Binney 1977; Rees \& Ostriker
1977; Silk 1977; White \& Rees 1978). Semi-analytic models of galaxy
formation in the CDM cosmology initially seemed to lend substance to
this view (White \& Frenk 1991, Kauffmann et al.\ 1993, Cole et al.\
1993, 2000; Somerville \& Primack 1999). However, these models assumed
a low value of the mean baryon density in the Universe, $\Omega_{\rm b}
\simeq 0.02$, which has now been ruled out by better
determinations. These indicate values twice as large (Spergel et al.\
2003, 2006, Cuoco et al.\ 2004, Sanchez et al.\ 2006). Benson et al.\
(2003) explicitly showed that in a cold dark matter universe with
such a high baryonic content, the cooling time argument alone is
insufficient to explain the characteristic size of present-day
galaxies, the sharpness of the break in the local galaxy luminosity
function and the relatively low observed number of very bright
galaxies.

Recent measurements of stellar mass reveal a substantial population of large
galaxies at $z=1$ and beyond (Drory et al.\ 2003, 2005, Pozzetti et al.\ 2003,
Kodama et al.\ 2004, Fontana et al.\ 2004, Bundy et al.\ 2005). Semi-analytic
models that match the luminosity function at $z=0$ reasonably well often
predict far fewer large galaxies at high redshift than is observed (Baugh et
al.\ 1998, 2004; Somerville, Primack \& Faber 2001); on the other hand, models
that give a better match to the abundance of high redshift massive
galaxies have typically failed to give good matches to the shape
of the luminsoity function locally (Kauffmann et al.\ 1999a, 1999b). In spite
of large uncertainties, observations also suggest that the star formation rate
in massive galaxies was larger at high redshift than it is in the local Universe
(Cowie et al.\ 1996, Juneau et al.\ 2005). These observations are contrary to
naive expectations for the hierarchical growth of structure in a CDM universe in
which large halos form late by the coalescence of smaller ones. Observational
astronomers have highlighted this apparent contradiction by coining phrases such
as ``cosmic downsizing'' (Cowie et al.\ 1996) or ``anti-hierarchical galaxy
formation.''

In this paper, we consider not only the conundrum of the galaxy
luminosity function and its evolution, but also two closely related
problems. The first is the observation that the most massive galaxies
today tend to be old and red, not young and blue as naively expected
in a hierarchical model. The second is the apparent absence of cooling
flows at the centres of rich clusters where the cooling times are
sufficiently short that large amounts of gas ought to be cooling but
very little is observed to do so (Tamura et al. 2001, Peterson et
al. 2003). In simple galaxy formation models it is this seemingly
missing cooling gas that results in the formation of the unobserved
population of extremely massive, blue galaxies. These are all facets
of the same problem and, in this paper, we argue that they all stem
from the neglect of a key phenomenon whose importance for galaxy
formation has only recently been recognised: the impact of the
formation of supermassive black holes at the centres of
galaxies. Granato et al.\ (2004) and Croton et al.\ (2006) have
proposed models of galaxy formation that include this phenomenon. The
work of Croton et al.\ in particular explicitly shows how these three
intertwined problems can be solved in the CDM cosmology by taking into
account the heating of gas in massive halos by energy injected by
active galactic nuclei. De Lucia et al.\ (2005) show further how this
same model naturally accounts for the apparently anti-hierarchical
nature of the star formation history of massive ellipticals. Our work
has many aspects in common with that of Croton et al.\ but our model
is different from theirs (particularly the treatment of feedback from
AGN) and, in this paper, we focus on its application to the high
redshift Universe
\footnote{While Croton et al., Springel et al.\ (2005a) 
and De Lucia et al.\ (2005) present a number
of results on the high redshift Universe, they do not focus on the evolution
of the mass and luminosity functions.}.

The shape of the mass function of dark matter halos is very different from the
shape of the galaxy luminosity function (see Fig.~1 in Benson et al.\ 2003). In
its most basic form, a theory of galaxy formation may be seen as a set of
physical processes that transform the halo mass function into the galaxy
luminosity function, that is, as a model for populating halos of different mass
with galaxies. The most powerful methodology to achieve this is semi-analytic
modelling, a technique that has revolutionised our way of understanding galaxy
formation. The key components are a background cosmology (e.g. the $\Lambda$CDM
model) which determines the formation and merger histories of dark matter halos;
the cooling of gas in halos; the formation of stars from cooled gas and feedback
processes that couple gas heating to the efficiency of star formation. In their
15 years of existence, semi-analytic models have met with many successes
(e.g. White \& Frenk 1991, Kauffmann et al.\ 1993, 1999a, 1999b, Cole et al.\
1994, 2000; Somerville \& Primack\ 1999, Somerville et al. 20001, Menci et al.\
2005, Baugh et al.\ 2005; and references therein). Yet, a number of fundamental
challenges remain, notably the problems just discussed.

The Durham semi-analytic model laid out in Cole et al.\ (2000; see
also Benson et al. 2002 and Baugh et al.\ 2005) was applied to the
problem of the galaxy luminosity function by Benson et al.\ (2003),
adopting $\Omega_{\rm b} = 0.045$ consistent with recent determinations. If
no feedback processes are included, the halo mass function is much
steeper at the low mass end than the galaxy luminosity function is at
the low luminosity end. In Benson et al. (2003) we showed that
plausible feedback processes (photoionization of pre-galactic gas at
high redshift and energy injection from supernovae and stellar winds)
would make galaxy formation sufficiently inefficient in small halos to
account for the relative paucity of low luminosity galaxies. However,
solving the faint-end problem in this way gave rise to a problem at
the bright end: the reheated leftover gas would eventually cool in
massive halos giving rise to an excessive number of bright
galaxies. We considered two plausible processes that could, in
principle, prevent this from happening: heat conduction near the
centre of massive halos and superwinds that would eject gas from these
halos. Unfortunately, neither of these solutions proved completely
satisfactory. To be effective, conduction would need to be implausibly
efficient, while supernovae do not generate enough energy to power the
required superwinds. We concluded that solving the bright-end problem
would involve tapping the energy released by accretion onto black
holes.

The mechanism that prevents the formation of extremely bright galaxies in
massive halos is almost certainly the same mechanism that prevents cooling flows
in clusters. An idea that has gained popularity in recent years is that energy
injected from AGN in the form of radio jets is responsible for switching off
cooling at the centre of massive halos (Quilis et al.\ 2001; Bruggen \& Kaiser
2002; Churazov et al.\ 2002; Dalla Vecchia et al.\ 2004, Sijacki \& Springel
2006). At the same time, a number of recent papers have considered the impact of
AGN energy on galaxy formation (e.g. Silk \& Rees 1998; Kauffmann \& Haehnelt
2000; Granato et al.\ 2004; Monaco \& Fontanot 2005, Cattaneo et al\ 2005;
Croton et al.\ 2006, De Lucia et al. \ 2005) although only Croton et al.\ have
calculated the resulting galaxy luminosity function\footnote{Granato et
al. (2004) also computed a luminosity function, but this was for high redshifts,
and only included the spheroidal component of galaxies.}. As emphasised by them
and also previously, amongst others, by White \& Frenk (1991), Benson et al.\
(2003), and Binney (2004), the sharp break at the bright end of the luminosity
function suggests a preferred physical scale in the problem. Such a scale arises
naturally from the interplay between two timescales: the timescale for a parcel
of gas to cool and the timescale for it to free fall to the centre of the
halo. For halo masses below $\sim 2\times 10^{11} \hMsol$, cooling is faster
than the free-fall rate, but in larger halos the cooling timescale exceeds the
sound crossing time and the cooling material is always close to hydrostatic
equilibrium. The transition between the rapidly cooling and the hydrostatic
regimes provides the physical basis for understanding the sharp break in the
galaxy luminosity function.  Binney (2004) argued that heat input from a central
source can only effectively prevent cooling in the hydrostatic case and Croton
et al.\ explicitly constructed a model demonstrating how this process can
explain the galaxy luminosity function.

In this paper, we include AGN feedback in the Durham semi-analytic galaxy
formation model, {\sc galform}, using the physical model just described. The
motivation for our treatment is similar to that of Croton et al.\ (2006), but
our implementation differs in key respects. In particular, Croton et al.\
parametrised the injection of AGN energy as a semi-empirical function of the
black hole and dark halo masses.  In contrast, our model assumes that the AGN
energy injection is determined by a self-regulating feedback loop. The black
hole mass only enters through the Eddington limit on the available power. The
power produced by the AGN does not depend on the mass of the dark matter halo
explicitly since this only enters through the cooling luminosity of the cooling
flow.  Like Croton et al., we implement our semi-analytic model on the
Millennium simulation of the growth of dark matter structure in the $\Lambda$CDM
cosmology recently carried out by the Virgo Consortium (Springel et al.\
2005a). In this way, we generate an alternative galaxy catalogue to that of
Croton et al. which is available at \website.

Our feedback prescription results in a model that provides a good description of
many observed properties of the local galaxy population. In particular, it
produces a good match to the luminosity function and the distribution of
colours. However, the successes of this model go beyond the properties of the
local Universe. We find that the model also successfully explains the evolution
of the galaxy stellar mass function out to $z=2$ and greater and that it
provides a natural explanation for the ``anti-hierarchical'' behaviour
championed by observational astronomers.

The model presented here differs in important ways from that of Baugh et al.\
(2005) who calculated a model without AGN feedback and found that, in that case,
a top-heavy stellar initial mass function (IMF) in bursts seemed to be required
in order to reproduce the observed numbers of ``sub-millimeter'' and
``Lyman-break'' galaxies. Using this model, Nagashima et al.\ (2005a,b) argued
that the observed chemical enrichment of the intracluster gas and of elliptical
galaxies is consistent with such a top-heavy IMF. In this paper, we adopt a
standard IMF throughout, but in future papers we plan to investigate the impact
of AGN on these issues. 

The remainder of the paper is structured as follows. In \S2, we outline the
differences between the present model and earlier versions of {\sc galform}. In
\S3, we briefly describe how we have adjusted the parameters of the new model
and present an overview of the comparison with galaxy properties in the local
Universe. In \S4, we fix this parametrisation and examine the predicted
properties of galaxies at higher redshift. In \S5, we turn our attention to the
volume averaged star formation history of the Universe, focusing in particular
on how this is apportioned amongst galaxies of different mass.  We summarise our
results in \S6.  Throughout the paper, we adopt the cosmological model assumed
in the Millennium simulation: a flat universe in which the density of cold dark
matter, baryons and dark energy (in units of the critical density), 
 have the values $\Omega_{\rm
b}=0.045$, $\Omega_{\rm m}=\Omega_{\rm CDM}+\Omega_{\rm b}=0.25$ and
$\Omega_\Lambda =0.75$ respectively, the Hubble parameter is $H_0=73\,
\hbox{km}\,\hbox{s}^{-1}\hbox{Mpc}^{-1}$, the power spectrum normalisation is 
$\sigma_8=0.9$ and the primordial spectral index is $n=1$. (We quote our
results in terms of the Hubble variable
$h=H_0/100\,\hbox{km}\,\hbox{s}^{-1}\hbox{Mpc}^{-1}$.)

\section{A new galaxy formation model}

The basic model that we use, {\sc galform}, is described in detail by Cole et
al. (2000) and Benson et al.\ (2003), and follows the fundamental principles
outlined by White \& Frenk (1991).  We have extended the model to track
explicitly the evolution of AGN and the impact of the energy emitted by AGN on
the cooling of gas in the host halo.  The revised model also incorporates a
better treatment of the gas reheated by star formation activity, allowing this
material to mix with an existing hot gas atmosphere and participate in any
ongoing cooling.

The starting point for our new model is the implementation described
by Benson et al. (2003) (not the version by Baugh et al. 2005 in which
a top-heavy stellar initial mass function was assumed in
starbursts). The substantial changes in parameters relative to that
model are an increase in the strength of feedback from supernovae (see
\S\ref{sec:local}), a slightly longer timescale for star formation in 
galactic disks ($\epsilon_* = 0.0029$) and an increase in the dynamical friction
timescale\footnote{The reader is referred to Cole et al. (2000) for
equations defining the model parameters discussed in the text. The
parameter $f_{\rm df}$ is defined in eqn. 4.16.}, $f_{\rm df}=1.5$. We
intend to present a full examination of the effects of parameter
variations on our model in a future paper. We adopt a Kennicutt (1983) IMF
with no correction for brown dwarf stars (ie., $\Upsilon=1$
in the language of Cole et al. 2000).

Our model follows Benson et al. (2003) in including a simple
prescription for the effects of the reionization of the Universe on
the formation of galaxies. In this approximation, we assume that gas
cooling is completely suppressed in dark matter halos with virial
velocities below $50\,\kms$ at redshifts below $z=6$.  {This treatment of
photoheating is a simple parametrisation of the results of the
detailed model calculated by Benson et al. (2002). These authors
computed the thermal history of the IGM and calculated the filtering
mass defined by Gnedin (2000) as the mass of the halo which, because
of photoheating, accretes only 50\% of the gas it would have accreted
from a cold IGM. Benson et al. (2002) find that the virial velocity
corresponding to the filtering mass lies between 50 and $60\,\kms$
between redshifts 0 and 6 and drops rapidly for higher redshifts. Our
simple model thus captures the overall behaviour of the more detailed
model and, as shown by Benson et al. (2003), it gives very similar
results.}

Besides these parameter changes the main enhancements in our current 
calculation relative to Benson et al. (2003) are:

\begin{enumerate}

\item {\bf Formation and growth of black holes.} Our model of black
hole formation is described in detail by Malbon et al. (2006), where we explore 
the properties of AGN emission and its evolution. It is based on the model of
Kauffmann \& Haehnelt (2000). 
In brief, black holes are assumed to grow through gas 
accretion triggered by galaxy mergers and disk instabilities as well 
as through mergers with other black holes. The growth rate is set by an
efficiency parameter, $\Fbh$, defined as the ratio of the gas mass
accreted onto the black hole to that formed into stars during a 
starburst induced either by a merger or by a disk instability.
In addition, we allow for black hole growth during
the feedback phase (see below). The results below correspond to $\Fbh=0.5\%$. 
This value was chosen to match the normalisation of the observed relation 
between black hole mass and galaxy bulge mass.

\item {\bf Disk instability.} This process was briefly discussed by Mo, Mao \&
White (1998) and by Cole et al. (2000). If a disk becomes sufficiently massive
that its self-gravity is dominant, then it is unstable to small perturbations by
minor satellites or dark matter substructures.  We find that this process plays
an important role in the growth of black holes in the Millennium
simulation in which low mass companions are
not resolved. We follow Cole et al.\ (2000) 
and use as the stability criterion the quantity
\begin{equation}
\epsilon=\frac{V_{\rm max}}{({\rm G} M_{\rm disk}/r_{\rm disk})^{1/2}},
\label{eq:dstab}
\end{equation}
where $V_{\rm max}$ is the circular velocity at the disk half-mass radius,
$r_{\rm disk}$, and $M_{\rm
disk}$ is the disk mass.  
If, at any timestep, $\epsilon$ is less than $\dstab$, the disk is
considered to be unstable. The entire mass of the disk is then transferred to 
the galaxy bulge, with any gas present assumed to undergo a starburst.

When a bulge is formed in this way, we compute its size using energy
conservation arguments as described by Cole et al. (2000).  
{We have, however, made an improvement in the way in which we
determine the size of a disk that becomes unstable.  Cole et
al. (2000) assumed that disks shrink to a size at which they become
rotationally supported and then estimated their total energy at this
size. We find that this method can occasionally produce disks of very
small size and very high circular velocity (several times that of the
host halo). As a result, spheroids formed from such disks are also
unrealistically small.  These disks have low angular momentum and have
been forced to become almost entirely self-gravitating in order to be
rotationally supported. They are therefore very unstable, with
$\epsilon$ well below $\dstab$. A more realistic approach is to assume
that the disk shrinks only until it first becomes unstable (as judged
by the above criteria) at which time a spheroid is formed.  This
ensures that unstable disks have reasonable sizes and circular
velocities. Therefore, for any disk deemed to be unstable, we find the
largest radius at which the disk would first become unstable and compute its
energy based on that radius.  }

As described above, when a disk becomes unstable, a fraction $\Fbh$ of
the gas goes into feeding the central black hole. We consider values
of $\dstab\sim1$ (Efstathiou, Lake \& Negroponte 1982).

\item {\bf Improved cooling calculation.}
To calculate the amount of gas that cools in a halo, Cole et
al. (2000) chopped the merger trees of halos up into branches defined
by the time it takes for a halo to double its mass. The propagation of
the cooling radius (defined as the radius within which the cooling
time of the gas equals the current age of the halo) is tracked during
these segments, but gas accreted from the intergalactic medium or
through infalling halos during the lifetime of the halo, or gas
returned to the halo by stellar feedback, was not allowed to cool
until the start of the next segment.

We have improved our cooling calculation by allowing for cooling of
material accreted by the halo during the current segment as well as
material reheated from galaxies by stellar feedback.  This new
calculation explicitly transfers reheated gas back to the reservoir of
hot gas on a timescale comparable to the halo's dynamical timescale.
We define $\tau_{\rm reheat} = \tau_{\rm dyn}/\arh$ and, over a
timestep $\Delta t$, increment the mass available for cooling by
$\Delta m = m_{\rm reheat} \Delta t / \tau_{\rm reheat}$, where we
have included a parameter $\arh$ to modulate this timescale,
anticipating $\arh\sim1$.

The model implicitly assumes that the gas in a halo is shock-heated
during collapse to the virial temperature before it begins to
cool. Birnboim \& Dekel (2003)and Kere\v{s} et al. (2005) have argued
that this is not realistic for galactic-sized halos in which gas
cooling may be so rapid that an initial shock may not form (see also
Navarro \& White 1993, Kay et al. 2000). As White \& Frenk (1991)
pointed out and Croton et al.\ (2006) have discussed in detail, the
propagation of the nominal cooling radius is so fast in these halos
(much faster than the halo dynamical time), that the supply of gas
from the halo is limited by the free-fall time and the halo's gas
accretion rate. The cooling time is irrelevant in small halos.

It is useful to distinguish between cooling that occurs on a free-fall timescale
and cooling that occurs from a quasistatic hot atmosphere. Following the
methodology of Cole et al. (2000), we compute the evolution of the cooling
radius and the free-fall radius as a function of halo age.  If the free-fall
radius is significantly smaller than the cooling radius, gas is accreted on a
free-fall timescale. Conversely, if the free-fall radius is significantly larger
than the cooling radius, cooling occurs in a quasi-hydrostatic cooling flow.

\item {\bf AGN Feedback.} We assume that AGN feedback is effective
only in halos undergoing quasi-hydrostatic cooling. Only in this
situation do we expect energy input from a central energy source to be
able to stabilise the flow and regulate the cooling rate.  To
establish which halos are susceptible to this kind of feedback, we
compare the cooling time at the cooling radius ($t_{\rm cool}(r_{\rm
cool})$, which is equal to the age of the halo) with the free-fall
time at this radius. If
\begin{equation}
{t_{\rm cool}(r_{\rm cool}) > \acool t_{\rm ff}(r_{\rm cool})}
\label{eq:acool}
\end{equation}
(where $\acool\sim1$ is an adjustable tuning parameter) we examine whether the central AGN
is able to inject sufficient power to offset the energy being radiated away in
the cooling flow. If the available AGN power is greater than the cooling
luminosity, we assume that the cooling flow is indeed quenched. 
 We parametrise 
the available AGN power as a fraction, $\ebh$, of the Eddington luminosity of the central
galaxy's black hole so that a halo is prevented from cooling if:
\begin{equation}
L_{\rm cool} < \ebh L_{\rm Edd}.
\end{equation}

Croton et al.\ (2006) also assumed that AGN feedback operates in
quasi-hydrostatically cooling halos (their ``radio'' mode).  Unlike their scheme,
however, our model for AGN feedback does not assume a phenomenological dependence of
the strength of the feedback on the gas temperature or black hole mass. Instead,
we simply assume that the flow will adjust itself so as to balance heating and
cooling whenever the Eddington luminosity of the black hole is sufficiently
large. A model is deemed acceptable if $\acool\sim1$ and $\ebh\lsim1.0$. 
{Note that although the Eddington luminosity provides a useful
measure of the energy scale, the jet power can, in principle, exceed
this limit since the outflow is strongly beamed.  In practise,
however, we find that it is unnecessary to make $\ebh$ this large.

During this feedback phase, the mass of the black hole grows as a
result of accretion from the quiescent inflow of gas cooling from 
the halo's hot atmosphere. We assume that jets are efficiently powered
with the black hole mass growing as $\dot{m_{\rm bh}}=L_{\rm cool}/0.2c^2$
(ie., the jet extracts half the available accretion energy in the 
case of a rapidly spinning black hole).  This growth channel, however,
is relatively unimportant until late times. In our model, black holes
gain most of their mass when they accrete gas funnelled to the
galactic centre during disk instabilities. Adopting a lower accretion 
efficiencies results in a larger contribution to the black hole mass 
density from this form of feedback but has very little effect on 
galaxy properties.
}

\item {\bf N-body merger trees.} We have implemented our new model on
merger trees constructed from the Millennium simulation using similar
techniques to those described by Helly et al.\ (2003). The Millennium
simulation, carried out by the Virgo consortium and described by
Springel et al.\ (2005a), followed the evolution of 10 billion dark
matter particles in a cubic volume of side $500\hmpc$. The particle
mass is $8.6 \times 10^8\hMsol$ so that about 20 million halos with
more than 20 particles form in the simulation. We take care to treat
halos close to the resolution limit (and those affected by the halo
deblending algorithm) correctly and, in particular, we ensure that
their chemical history is consistent with higher resolution trees.  
The smallest resolved halos
typically host galaxies with luminosities less than around 0.03$L_*$.
Our sample of model galaxies is therefore complete for luminosities
brighter than this value.

The properties of our merger trees have been described by Harker et
al.\ (2006). These were constructed independently of the trees
presented by Springel et al.\ (2005a) which were used by Croton et al
(2006). The two sets of trees differ in several respects most notably
in the criteria used for separating halos spuriously linked by the
group finder, in the definition of independent halos, and in the way
in which the descendants of halos are identified at subsequent times.

\end{enumerate}

\section{The Local Universe}
\label{sec:local}

\begin{figure}
\includegraphics[width=8.6cm]{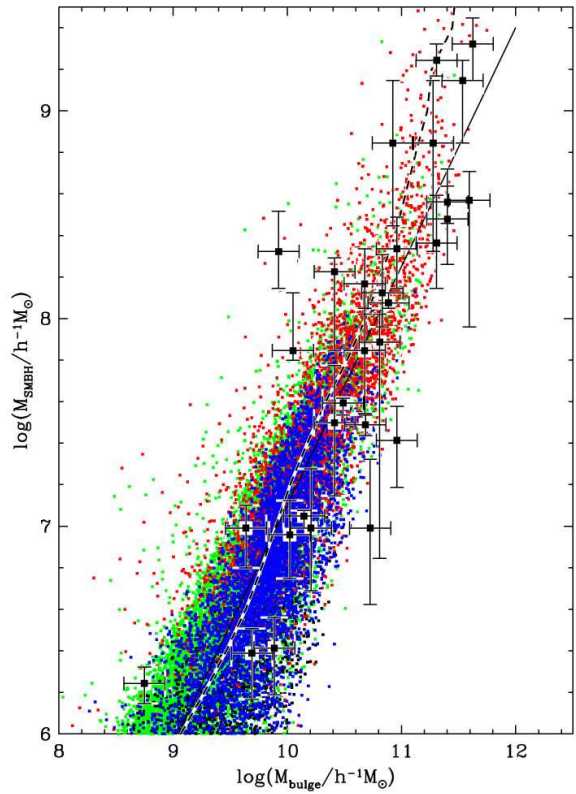}
\caption{The black hole {\it vs.} galaxy bulge mass relation at $z=0$. The
colour dots show the relation obtained using one tenth of the
Millennium simulation. The colours distinguish galaxies according to
whether they are satellites (green points) or central galaxies
of halo of masses $<3\times 10^{11} h^{-1}M_\odot$ (black points), $3\times 10^{11}$--$10^{12} h^{-1}M_\odot$ (blue points) or $>10^{12} h^{-1}M_\odot$ (red points). The points with error bars are local data 
from Haering \& Rix (2004); the black line gives the best fitting
relation to these data.}
\label{fig:bhmass}
\end{figure}

\begin{figure} \includegraphics[width=8.6cm]{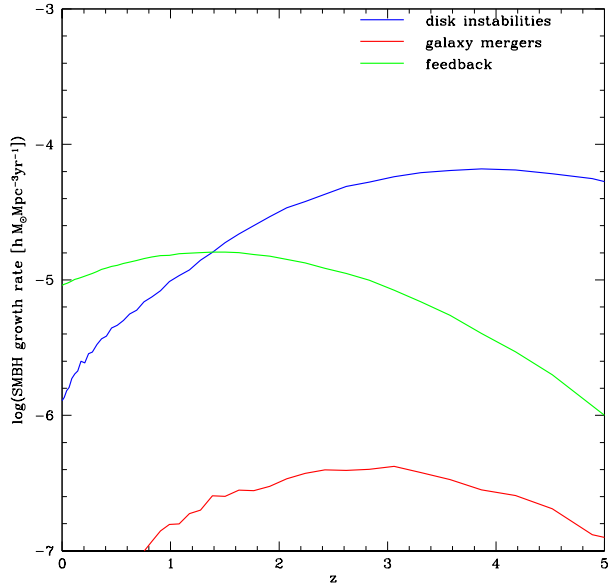}
\caption{ {The
contribution of different processes to the volume averaged rate of
growth of black hole mass as a function of redshift.  The colour lines
illustrate the contribution to the black hole mass growth rate from:
galaxy mergers (red); disk instabilities (green); accretion associated
with AGN feedback in quasi-hydrostatic flows (blue).  At high
redshift, the growth of black holes is dominated by instabilities in
the rapidly forming disks. At lower redshifts, it is dominated by
accretion from quasi-hydrostatic cooling in massive halos.}}
\label{fig:bhgrowth} 
\end{figure}

\begin{figure}
\includegraphics[width=8.6cm]{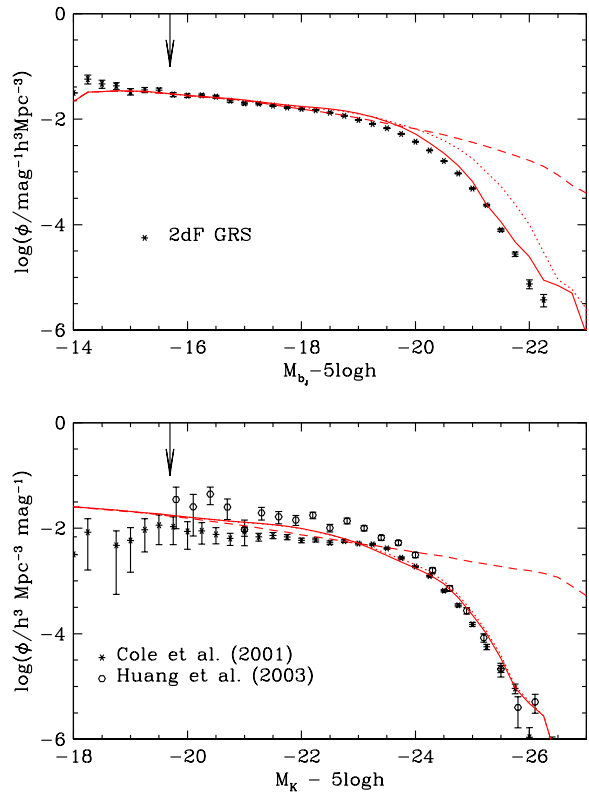}
\caption{The luminosity function of galaxies in the local Universe. The upper
panel compares the model b$_J$-band luminosity function (red lines)
with the observational determination from the 2dF galaxy redshift
survey by Norberg et al. (2002). Here and in the panel below the
dotted line shows the model prediction without dust obscuration and
the solid line the prediction taking obscuration into account, 
while the dashed lines show models in which feedback from AGN has been
switched off. The 
lower panel compares the K-band luminosity function in the model to
the observational determinations by Cole et al. (2001) and Huang et
al. (2003). Arrows indicate the approximate magnitude faintwards of
which of sample of model galaxies becomes incomplete due to the
limited mass resolution of the Millennium simulation.}
\label{fig:z0lf}
\end{figure}

\begin{figure}
\includegraphics[width=85mm]{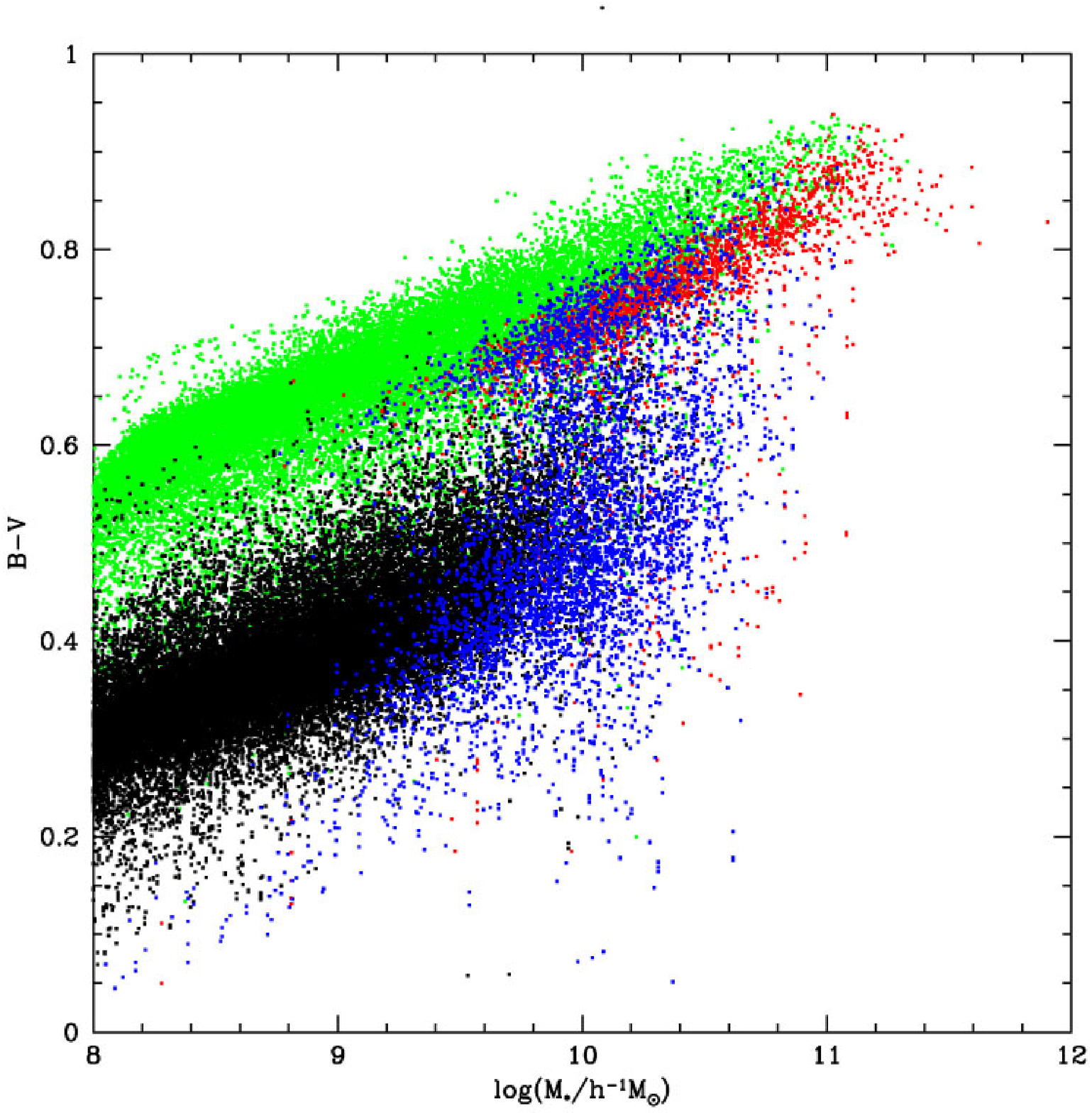}
\caption
{The colour distribution of model galaxies from one tenth of the
Millennium simulation. The graph shows 
the B\-V colour as a function of stellar mass in a galaxy. Note the
clearly bimodal distribution of galaxy colours, and the transition to
a unimodal colour distribution which occurs at $M_*\sim 10^{10.5}
\hMsol$. The points are colour coded as in Fig.~\ref{fig:bhmass},
i.e. satellites (green points) and central galaxies 
of halo of masses $<3\times 10^{11} h^{-1}M_\odot$ (black points),
$3\times 10^{11}$--$10^{12} h^{-1}M_\odot$ (blue points) or $>10^{12}
h^{-1}M_\odot$ (red points).} 
\label{fig:z0colours}
\end{figure}

In keeping with the philosophy advocated by the Durham semi-analytic
modelling group, we require that any acceptable model should match
basic properties of the local galaxy population, particularly the
galaxy luminosity function. We fix the adjustable parameters of our
model to achieve this. In this application, we also require that the
model should match the observed relationship between black hole mass
and galaxy bulge mass. This is important since we find that AGN
feedback plays an important role in shaping the local luminosity
function.

The recent observational determination of the present-day black hole-bulge mass
relation by Haering \& Rix (2004) is illustrated in
Fig.~\ref{fig:bhmass}. The symbols with error bars show the
observational data, with the best-fit from that paper shown as a black
line. Our model was adjusted to match this relation by varying both
the level at which disks become unstable ($\dstab$) and the fraction
of the cold gas mass that is fed to the black hole during a starburst
($\Fbh$). Some iteration was required to obtain an acceptable model
since the growth of black holes and bulges depends on other parameters
of the model. Setting $\dstab=0.8$ and $\Fbh=0.5\%$ achieves the match
shown in Fig.~\ref{fig:bhmass}. This value of $\dstab$ is reasonably close to the
value of 1.1 for the stability threshold found by Efstathiou et al.
(1982) (who, however, used slightly different definitions of the
quantities that enter into eqn.~\ref{eq:dstab}), but results in a somewhat 
larger fraction of disks becoming unstable. It is interesting to note that
the model black hole-bulge mass relation steepens at high bulge masses. This 
is a result of the growth of the black hole during the feedback phase.
This steepening ensures that the Eddington luminosity of the black hole
is sufficient to quench cooling even in the most massive clusters.

{The relative importance of the different channels that lead
to the growth of black hole mass is illustrated in
Fig.~\ref{fig:bhgrowth}. The colour lines represent different black
hole growth mechanisms. Because of the limited mass resolution of the
Millennium simulation, we can only follow mergers of relatively
massive satellites (mass ratios greater than 0.1). Central gas flows
triggered by mergers therefore play a relatively minor role in the
build-up of black holes. At high redshift, black hole growth  is
dominated by instabilities in rapidly forming ``disks'', while at
lower redshift it is dominated by the mass accreted during the
feedback cycle that quenches cooling in massive halos.}

Accounting for the sharp cutoff at the bright end of the luminosity function
has proved challenging for previous semi-analytic models, particularly
if the baryon density $\Omega_{\rm b}$ is as large as implied by
modern determinations (Benson et al. 2003). In our new model, four key
parameters control the shape of the luminosity function: $\ahot$,
$\vhot$, $\acool$ and $\arh$. The first two are the parameters
introduced by Cole et al. (2000) to describe the conventional feedback
produced by energy injected by supernovae and stellar winds. They
essentially determine the faint end slope of the luminosity
function. 

The new parameter associated
with AGN feedback, $\acool$ (which controls when a halo is considered
to be cooling quasi-hydrostatically as defined in
eqn.~\ref{eq:acool}), determines the position of the exponential break
in the galaxy luminosity function. Increasing $\acool$ shifts the
break to lower luminosities. Finally, the overall normalisation of the
luminosity function is strongly affected by the parameter $\arh$ that
determines the timescale on which reheated disk gas is added to the
hot gas reservoir.

The models presented here have $\ebh=0.5$, $\ahot=3.2$, $\vhot=485\,\kms$,
$\acool=0.58$ and $\arh=0.92$. For this choice, the model gives an
excellent match to the local $\rm b_{\rm J}$ and K-band galaxy luminosity functions,
as shown in Fig.~\ref{fig:z0lf}. (The dashed lines in this figure show
the luminosity function when feedback from AGN is switched off. The
match to the observational data is then very poor, due in part to the
large value of $\vhot$. Benson et al. (2003) and Baugh et al. (2005)
were able to obtain better matches to the luminosity function---albeit
not as good as in our model including AGN feedback---through the
inclusion of alternative means of suppressing gas cooling in massive
halos.) 

{The values of $\ahot$ and $\vhot$ that we have adopted are
somewhat larger than the values used in our previous work and imply
that stellar feedback is extremely strong, especially in small
galaxies. As a result, the energy involved in feedback (i.e. the
energy required to heat cold disk gas to the halo virial temperature)
is equal to the total energy produced by supernovae in galaxies with
$v_{\hbox{circ}}=200\,\kms$ and exceeds the available supernovae
energy by $(v_{\hbox{circ}}/200\,\kms)^{-1.2}$ in lower mass galaxies.
It should be noted, however, that our model does not make use of any
of the available AGN energy to eject material from the galaxy
itself. Numerical simulations of merging galaxies with central
supermassive black holes indicate that this can be a significant
source of additional feedback even though the outflow is strongly
beamed (Di Matteo et al.\ 2005, Springel et al.\ 2005b).  We also
require that the material be heated to the virial temperature
of the halo.  Both of these assumptions are likely to lead to an 
overestimate of the energy injection that is required. In addition, the
values of these feedback parameters should be treated with caution
since our model of star formation is undoubtedly simplified. We intend
to present a more detailed investigation of the star formation and
cooling model in future work.  }

Fig.~\ref{fig:z0lf} shows how AGN feedback can solve one of the two
problems highlighted in the Introduction: the absence of the very
bright galaxies that unquenched cooling flows would generate at the
centres of very massive halos. Fig.~\ref{fig:z0colours} shows how our
model solves the second problem: the red colours observed for the most
massive observed galaxies. The colour distribution in the model is
clearly bimodal (c.f. Menci et al. 2005) and shows a very well defined
red sequence which, at the red end, is populated by the most massive
galaxies. The transition in mean colour occurs at around a mass of
$\sim 2\times10^{10}\hMsol$.  Our model gives a qualitatively good match to
observational data (\ Kauffmann et al.\ 2003, Baldry et al.\ 2004). 

The colour distribution displayed in Fig.~\ref{fig:z0colours} is influenced
strongly both by AGN feedback and by the improvements to our cooling
model which now allows reheated disk gas to be added to a surrounding
hot reservoir during the lifetime of a halo. This tends to establish a
near steady state in which gas cooling is balanced by reheating. Below
a halo mass of $\sim 10^{12}\hMsol$, the colours of central galaxies
follow a well-defined blue locus. Satellite galaxies that have been
stripped of their hot gas reservoirs rapidly migrate to the red
sequence. For higher halo masses, the AGN feedback becomes important,
and central galaxies also evolve onto the red sequence.  This process
is responsible for the abrupt change in galaxy properties at a stellar mass of
$2\times10^{10}\hMsol$ discussed by Kauffmann et al. (2003).

\section{The Distant Universe}

\begin{figure}
\includegraphics[width=8.6cm]{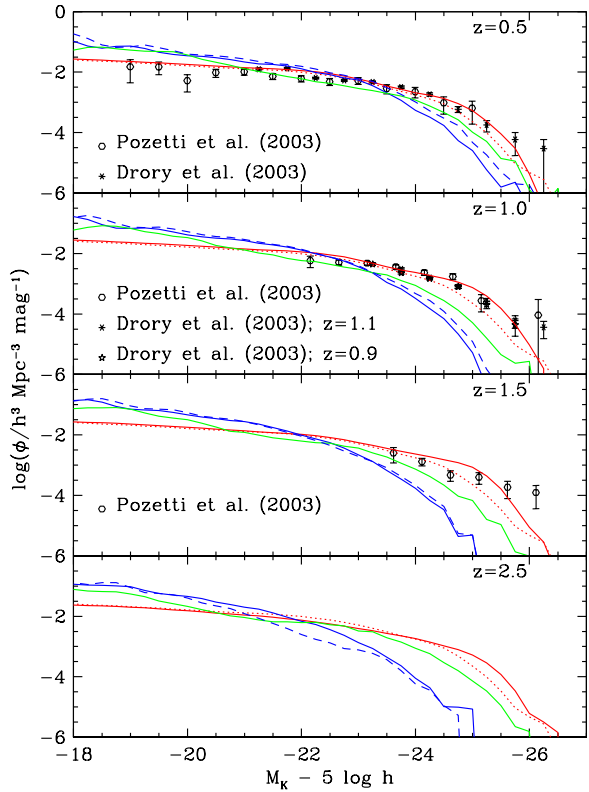}
\caption
{The evolution of the rest-frame K-band luminosity function. The
predictions of our new model at $z=0.5, 1.0$, $1.5$ and $2.5$ are shown as
red solid lines. The red dotted line in each panel corresponds to the model
luminosity function at $z=0$. The blue and green lines show, for
comparison, the rest-frame K-band luminosity function from the models
of Cole et al. (2000) and Baugh et al. (2005) respectively. The
observational data  come from the  
K20 survey of Pozzetti et 
al. (2003; circles, based on spectroscopic redshifts) and the MUNICS
survey of Drory et al. (2003; stars, based on photometric
redshifts). At $z=1.0$ we plot the data from Drory et al. $z=0.9$ and $z=1.1$ data
for comparison. 
}
\label{fig:klfevo}
\end{figure}

\begin{figure}
\includegraphics[width=8.6cm]{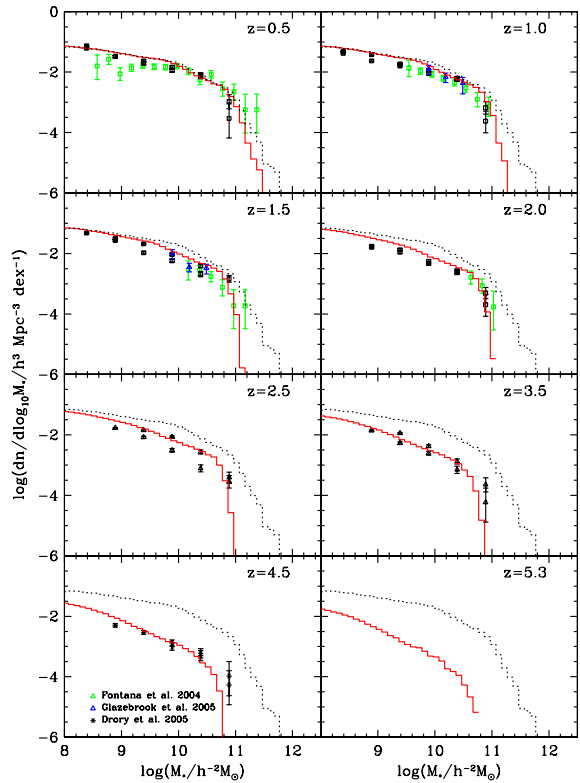}
\caption
{ The evolution of the stellar mass function. The model predictions
are shown by the red solid lines at the redshifts indicated in the
legend. The dotted line in each panel corresponds to the model stellar
mass function at $z=0$. The observational estimates from Drory et
al. (2005), based on photometric redshifts, are shown by symbols with
error bars. Data from the FORS deep field and the GOODS-S fields are
plotted separately and illustrate the variation arising from the small
volume surveyed.  
{We also show estimates of the stellar mass function
from spectroscopic surveys: Fontana et al.\ (2004, K20) and
Glazebrook et al.\ (2004, GDDS). In the region of overlap there is
good agreement between all three surveys.}}
\label{fig:starmfevo}
\end{figure}

\begin{figure}
\includegraphics[width=8.6cm]{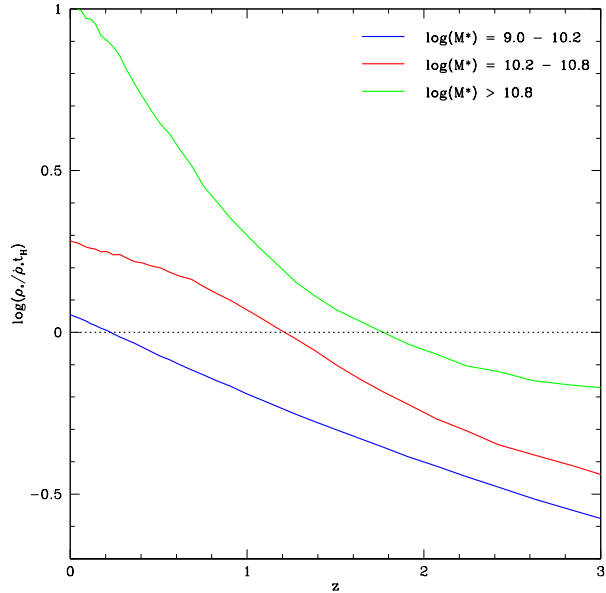}
\caption
{\bf{The evolution of the ratio of the past to present star formation rate 
for galaxies of different masses.  The colour lines show how the ratio of the 
past average star formation rate compares to the 
star formation rate at redshift $z$. Galaxies are
binned according to their stellar mass at the epoch, $z$, of
observation. The black dashed line divides the plot in to regions
in which the stellar mass of the galaxies is rapidly growing (below the line) 
and in which the galaxies have already largely formed (above the line).
Although this plot neglects young stars accreted through mergers, it 
indicates that higher mass galaxies have completed their star formation 
at higher redshifts than their low mass counter-parts.
}}
\label{fig:downsizing}
\end{figure}

\begin{figure}
\includegraphics[width=8.6cm]{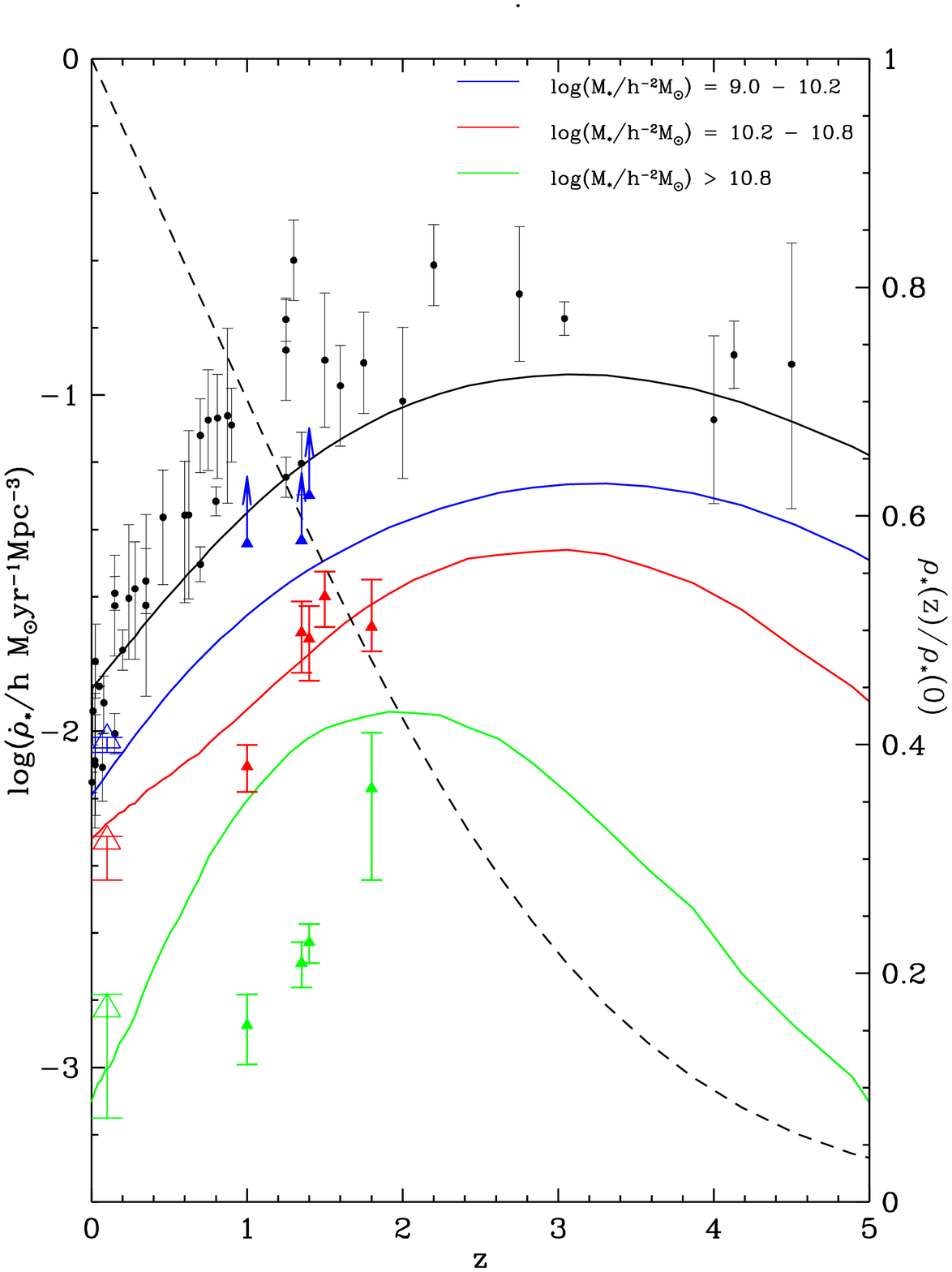}
\caption
{The evolution of the cosmic star formation rate per unit volume and
its dependence on galaxy mass. The black points show observational
determinations of the volume averaged star formation rate density as a
function of redshift taken from the compilation by Hopkins (2004). The
solid black line shows the evolution predicted by our galaxy formation
model. The dashed line shows the cumulative mass locked in stars by
redshift $z$ normalised to the value at the present day. The scale is
given at the right-hand side of the plot. The colour points and lines
show how the star 
formation rate is apportioned amongst galaxies of different mass. The
points show observational data from Juneau et al. (2005; filled
symbols) and Brinchmann et al. (2004; open symbols) while the lines
are the model predictions. Galaxies are 
binned according to their stellar mass at the epoch of
observation. For all masses, the star formation rate in the model increases from
$z=0$ to higher redshift and reaches a peak at $z=2$--3 for all
mass bins.
}
\label{fig:sfrd}
\end{figure}

Galaxy formation models in the $\Lambda$CDM cosmology have consistently had
difficulty in simultaneously accounting for the present day galaxy luminosity
function and for the number of apparently massive galaxies observed at redshifts
$z\sim 1$--$2$ (Pozzetti et al. 2003; Drory et al.\ 2003, 2005, Kodama et al.\
2004, Fontana et al.\ 2004). Thus, for example, the models of Kauffmann et al
(1999b) produced a large number of high redshift galaxies but they also produced
an excess of bright galaxies and a deficit of $L_{*}$ galaxies 
locally (Kauffmann et al 1999a). Conversely, the
models of Cole et al. (2000) and Baugh et al. (2004) gave a better match for the
local galaxy luminosity function but underpredicted the number of galaxies at
$z\sim 1$--$2$. At higher redshift still, $z\sim 2-5$, Baugh et al. (2005) have
shown that the abundance of submillimeter and Lyman-break galaxies can be
explained by the standard semi-analytic model of Benson et al. (2003) only if
the bursts of star formation triggered by mergers are assumed to have a
``top-heavy'' stellar initial mass function (IMF). In this paper we explore how
the inclusion of AGN feedback alters the predictions of the model for the
evolution of the stellar mass function in galaxies. We defer a detailed study of
the properties of the submillimeter and Lyman-break population to a later paper.

As in previous papers in this series, we have set the parameters of
the model to match a selection of global properties of the local
galaxy population. We now proceed to compare this model with the high
redshift Universe without any further adjustments to the
parameters. The most direct way to test whether the model predicts the
correct distribution of stellar mass in galaxies at high redshift is
to compare with observations of the rest-frame K-band luminosity
function. The K-band light is relatively insensitive to the
instantaneous star formation rate and thus provides a reasonable
tracer of the underlying stellar mass (Kauffmann \& Charlot
1998). Alternatively, the model predictions can be compared with the
result of attempts to estimate the stellar mass directly by combining
K-band data with photometry in bluer bands. This diagnostic uses
fuller information about the spectral energy distribution, but it has
the disadvantage that the inferred stellar masses are model dependent
and require assumptions for the stellar IMF and star formation
history. However, this technique can be applied to galaxies at high
redshift for which only photometric data and photometric redshifts are
available.

The comparison between our model predictions and the measured K-band
luminosity function at high redshift is shown in
Fig.~\ref{fig:klfevo}.  The measurements from the K20 survey (Pozzetti
et al. 2003) are at $z=0.5, 1.0$ and $1.5$ whereas the measurements
from the MUNICS survey (Drory et al. 2003) are at redshifts $z=0.5,
0.9$ and $1.1$. The model predicts a slight brightening of the
characteristic luminosity, $L_*$, with increasing redshift of very
similar size to the brightening seen in the real data. Both model and
data indicate that there is a population of galaxies at $z=1.5$ which
are luminous in the rest-frame near infrared and therefore contain
substantial stellar mass. This is the first time that a semi-analytic
model in the $\Lambda$CDM cosmology which gives a close match the
local galaxy luminosity function has provided such good agreement with
data of this kind. For example, we show the rest-frame K-band
luminosity functions from the models of Cole et al. (2000) and Baugh
et al. (2005) in Fig.~\ref{fig:klfevo}. These models reproduce the
$z=0$ K-band luminosity function reasonably well, but the luminosity
function evolves too fast with redshift and underpredicts the
abundance of galaxies bright in K at high redshifts\footnote{Other
attempts to reproduce the evolution of the K-band luminosity function
fail in a similar way---see the models plotted in Fig.~10 of Pozzetti et
al. (2003) for example.}. We discuss the physical reason for this
agreement in Section~6.

Testing the model at redshifts $z\gsim 1.5$ is complicated by the
difficulty of measuring the rest-frame K-band luminosity directly. At
these redshifts, the rest-frame K-band is shifted well outside the
near-IR ground-based observing window. In addition, spectroscopic
redshifts become progressively harder to determine. To try and
overcome these difficulties Drory et al.\ (2003, 2005) have applied
photometric techniques to determine both the redshift of the best
fitting template of the spectral energy distribution of a galaxy as
well as its total stellar mass.  Since the inference of the stellar
mass requires an assumption about the IMF, to compare the data with
our model, we have transformed the data to the Kennicutt (1983) IMF
assumed in our model (eg., we adopt a correction of $-0.3$ dex to
convert results given for the Salpeter IMF to the IMF assumed here,
Bell et al. 2003).  Drory et al.\ (2005) analysed two deep fields: the
FORS deep field, and the GOODS-S field. Their stellar mass functions
for the two fields, transformed to our IMF, are plotted separately in
Fig.~\ref{fig:starmfevo}. The stellar masses plotted exclude the mass
recycled to the interstellar medium. The two surveyed fields are small
and field-to-field variations are clearly noticeable.  A similar
approach has been employed by Fontana et al.\ (2004) and Glazebrook et
al.\ (2004) to determine stellar masses in the K20 and GDDS surveys
respectively for which spectroscopic redshifts are available. In the
region of overlap the spectroscopic studies give results results
consistent with the photometric estimate, as shown in the figure.

Using the photometric redshift technique, the evolution of the stellar
mass function can be traced back to $z=4.5$. Our model predictions are
shown by the solid line in Fig.~\ref{fig:starmfevo}, with the model
mass function at $z=0$ plotted as a dotted line for
comparison. The match of the model to the data is encouragingly good: 
both the overall normalisation and the position of
the break are well reproduced in the model even at the highest
redshifts probed. There is a hint that, at these redshifts, the model
may be underpredicting the number of galaxies in the brightest
bin. These galaxies are, of course, quite rare and furthermore, the
comparison in Fig.~\ref{fig:starmfevo} ignores the quite substantial
errors in the photometric redshift and mass-to-light ratio estimate. At the
highest redshift, the error in $M/L$ can be as large as 1 dex. This has
the effect of smoothing the sharp break seen in the model, and may account
for some of the apparent discrepancy (accurate modelling of the photometric 
mass estimator is required to test this). Our model thus predicts that
massive galaxies were already in place at redshift $z\simeq 4$ with a
stellar mass function which is very similar to that observed. We
discuss the reasons for this behaviour in Section~6.

{Since the model predicts that substantial stellar mass is
already in place in bright galaxies at high redshift, it is
instructive to compare the current star formation rate in these
galaxies with the past average star formation rate as a function of
galaxy mass. We make this comparison by plotting the ratio
$R=\rho_{*}/\dot\rho_{*}t_H(z)$ (where $\rho_{*}$ is the volume
density of stars in galaxies in a given mass range and $t_H(z)$ is the
Hubble time at redshift $z$) as a function of redshift for a range of
galaxy masses. This is shown in Fig.~\ref{fig:downsizing}, where blue,
red and green lines correspond to the stellar mass ranges
$10^9$--$10^{10.2}$, $10^{10.2}$--$10^{10.8}$ and
$>10^{10.8}\,h^{-1}M_\odot$ respectively.  The black dashed line at
$R=1$ divides the graph into two regions. Below the line, the global
average rate at which stars formed in the previous Hubble time is
smaller than the current rate of star formation at redshift z.  In a
globally averaged sense, these galaxies are still forming.  In
contrast, galaxies that lie significantly above this line have a
current star formation rate that will only add a small contribution to
the stars already in existence. As the plot illustrates, the most
massive galaxies cross this boundary at $z\sim2$, while the lowest
mass galaxies are only just crossing the boundary at the present-day.
This behaviour is often referred to as ``downsizing'' in the star
formation history of the Universe.}

\section{The cosmic star formation rate density}

The evolution of the volume averaged star formation rate density out to $z=5$ is
shown in Fig.~\ref{fig:sfrd}. The model prediction, shown by a solid black line,
is compared with a compilation of observational determinations by Hopkins (2004)
(converted to the Kennicutt (1983) IMF adopted here). The model provides a good
match to the data over most of the redshift range, except perhaps at the very
lowest redshifts, $z\lsim 0.3$, where the model overpredicts the star formation
rate by a about 20\%, and at $z\sim1$ where the model lies below the median of
the data by a similar factor. The overall effect is that the decline in the star
formation rate may be slightly weaker than suggested by the observations;
however, given the observational uncertainties (both random and systematic) the
model provides an adequate match. The dashed line gives the cumulative mass
locked up in stars by redshift $z$, normalised to the present value. Twenty
percent of the current stellar mass was in place by $z=3$ and half by $z=1.6$.

The colour lines show the contribution to the global star formation
rate from galaxies of different mass (at the time when they are
observed). We have split the model data into the same mass bins used
in the observational study of Juneau et al. (2005), the results of
which are also shown in the figure: $10^9$--$10^{10.2}$,
$10^{10.2}$--$10^{10.8}$ and $>10^{10.8}h^{-1}M_\odot$ (blue, red and green lines
respectively). Most of the star formation in the model takes place in
galaxies whose instantaneous stellar mass lies in the range
$(10^9$-$1.6\times 10^{10})\hMsol$, although at the present day, there
is a comparable contribution from galaxies with mass in the range
$(1.6$--$6.3 \times) 10^{10}\hMsol$. At all times, there
is a smaller contribution from galaxies in the most massive bin
plotted, $>6.3\times 10^{10}\hMsol$.

The time dependence of the star formation rate in all three stellar
mass bins has a similar shape: there is a broad peak at $z\sim 2$--3
with a strong decline below between $z=2$ and the present day.  The decline 
is particularly rapid in the highest mass bin and slower in the lowest mass
bin. Nevertheless, star formation activity in the most massive
galaxies peaks at a similar high redshift as in smaller galaxies. At first 
sight, this appears puzzling in the context of the hierarchical
nature of structure growth characteristic of the CDM cosmology. This
apparently paradoxical behaviour is, in part, the motivation for the 
terms ``downsizing'' and ``anti-hierarchical'' sometimes used to describe 
the observed history of star formation (and quasar) activity. 

There are two effects in the model that underlie the behaviour seen in
Figs. \ref{fig:downsizing} and~\ref{fig:sfrd}. 
Firstly, at a fixed halo mass, halos that collapse
at higher redshift have a lower ratio of cooling time to free-fall
time. This makes AGN feedback relatively less effective at higher redshift.
Since AGN feedback increases in importance with increasing halo mass, 
halos that collapse at high redshift are more likely to form
stars efficiently than those that collapse at low redshift. Secondly, 
this trend is enhanced because black holes are
smaller at earlier times. As a result, the few rare massive halos that
form at high redshift are able to make stars more efficiently (even if they are in the 
hydrostatic cooling regime) than 
their low redshift counterparts which have much larger black holes and 
thus more powerful AGN.

Our model prediction for the split of star formation rate density by
galaxy stellar mass can be tested against the data from the GDDS
survey of Juneau et al. (2005). This survey targeted faint galaxies in
the K-band at $z<2$, selecting according to stellar mass and star
formation rate, inferred from OII emission and the rest-frame far-UV
continuum. These data, together with lower redshift determinations
from Brinchmann et al.\ (2004) are shown as colour points in
Fig.~\ref{fig:sfrd}. The model agrees reasonably
well with the data in the range $1<z<2$, although the contribution
from the lowest mass bin is rather low at this redshift.
 We intend to refine the model by introducing a more detailed
 treatment of star formation and feedback in low mass halos and expect
 that this may reduce the remaining discrepancies seen in
 Fig.~\ref{fig:sfrd}.

\section{Discussion and conclusions}

We have presented a new model of galaxy formation in the $\Lambda$CDM
cosmology and implemented it in the Millennium simulation of Springel
et al. (2005a). Our galaxy formation model builds upon the hierarchical
model of Cole et al.\ (2000) by including a more complete treatment of
gas ejected from galaxies by stellar feedback and a new physical
process: feedback from AGN embedded in quasi-hydrostatic cooling
flows. 

The model parameters have been adjusted to produce a good match to the 
observed properties of local galaxies. Fig.~\ref{fig:z0lf} compares the
model with the local B and K band luminosity functions showing the level
of agreement that can be achieved. The colour distribution of the model
galaxies is shown in Fig.~\ref{fig:z0colours}. It is clearly bimodal,
with a transition in the typical colours of galaxies around a stellar
mass of $\sim 2\times10^{10}\hMsol$ as observed in the local Universe.

In this paper, however, we focus on the evolution of the stellar mass function
and star formation history predicted by the model.
A particularly interesting feature of our new model is the counterintuitive nature
of the star formation history that it predicts. Although the number density of
halos of mass $\gsim 10^{12} \hMsol$ is increasing towards low redshift (Mo \&
White 2002), star formation in them becomes increasingly less efficient and, as
a result, their contribution to the global star formation rate declines with
redshift. Such apparently ``anti-hierarchical'' behaviour is present even in the
early semi-analytic models of White \& Frenk (1991). However, in these models
the growth of halo mass was compensated for only by the increase in the cooling
time of the gas. With modern determinations of $\Omega_{\rm b}$, this effect is
insufficient to stop the formation of overly bright galaxies by cooling flows at
the centres of massive halos. The inclusion of AGN feedback solves this problem
and, of course, also explains why cooling gas is not observed in the amounts
inferred from cooling time arguments in the centres of galaxy clusters (Tamura et al. 2001, Peterson et al. 2002).

Two additional factors contribute to the relative decline in galaxy formation
activity in massive halos at low redshift. At a fixed halo mass, the ratio of
the cooling time of the gas to the free-fall time is shorter at high redshift
than at low redshift. Our model is based on the assumption that AGN feedback is
effective only during the quasi-hydrostatic cooling phase which today occurs in
halos of mass $M\gsim 3\times10^{11}\hMsol$. At high redshifts, the cooling
times in these halos were short enough to allow the gas to cool on the free-fall
timescale thus rendering any AGN feedback ineffective. In addition, since black
hole masses are lower at higher redshift and, in our model, the AGN feedback
is limited by the black hole's Eddington limit, the feedback is less effective
even in those halos that are already in the quasi-hydrostatic cooling
phase. These are the main reasons why our model predicts a larger number density
of massive galaxies at high redshift compared to earlier semi-analytic models
and also the reason why the most massive galaxies today completed their star
formation early and are now old and red.

In spite of the fact that our model agrees reasonably well with observational
determinations of the stellar mass function out to the highest redshift,
$z=4.5$, at which this function has been estimated, the model predicts that only
$\sim 20\%$ of the mass locked up in stars today was in place at
$z=3$. Similarly, the median redshift for star formation is
$z_{\rm med}=1.6$. These numbers are not very different from those predicted by many
previous semi-analytic models (e.g. Cole et al. 1994, 2000) which did not
include AGN feedback. This is partly because massive galaxies contain only a
relatively small fraction of the cosmic stellar population.

Although the two have been developed independently, the motivation for our model
of AGN feedback and the basic physical principles behind it are similar to those
discussed by Croton et al. (2006). However, there are a number of important
differences between these two models, most notably in the implementation of the
AGN physics, the cooling model used and the construction of the halo merger
trees from the Millennium simulation. While Croton et al. (and De Lucia et
al. 2005) presented extensive properties of their model at $z=0$ and discussed
only a few aspects of their evolution, in this paper we have focused on
properties at high redshift, particularly on the stellar mass function and star
formation rate as a function of redshift, which were not considered by Croton et
al.

More generally, the model presented in this paper and that of Croton et al.
should be regarded as two plausible implementations of the current understanding
of the physics of galaxy formation on the same underlying dark matter
distribution. Since the differences in the construction of the merger trees are
relatively minor (Helly et al., in preparation), the differences between the two
models reflect different choices for the way in which the various physical
processes are treated and parametrised. Both models give a similar match to the
luminosity function and colour distribution of galaxies in the local Universe
but they have, in fact, been constructed with this aim in mind. More interesting
are the similarities and differences between the two models for other properties
of local galaxies and, more importantly, properties of galaxies at high
redshift. These similarities and differences will be explored in future work
and will provide a good guide to the level of theoretical uncertainty in studies
of this kind.

In summary, we have presented, and implemented in the Millennium
simulation, a new model of galaxy formation in which AGN feedback is
responsible for the absence of cooling flows in rich clusters, for the
cutoff at the bright end of the galaxy luminosity function and for the
number density and properties of the most massive galaxies at all
redshifts. Galaxy catalogues constructed from it are 
available via the web at \website.

\section*{Acknowledgements}

RGB thanks PPARC for the support of senior fellowship. CMB and AJB
acknowledge the receipt of a Royal Society URF.  This work was
supported by PPARC rolling grant PP/C501568/1.  We thank Andrea
Cattaneo, James Binney and Darren Croton, and especially Simon White,
for valuable comments and discussions that have helped shape this
work.

\end{document}